\documentclass[aps,10pt,prl,twocolumn,showpacs,preprintnumbers,amsmath,amssymb,nofootinbib,floatfix]{revtex4-1}

\usepackage{amsmath}
\usepackage{graphicx}
\usepackage{multirow} 
\usepackage{array} 
\usepackage{longtable} 
\usepackage{bigstrut} 
\usepackage{verbatim}

\newcommand{\Aa}{\alpha}
\newcommand{\Bb}{\beta}
\newcommand{\Cc}{\gamma}
\newcommand{\Dd}{\delta}
\newcommand{\Ee}{\epsilon}
\newcommand{\Ff}{\zeta}
\newcommand{\Gg}{\eta}
\newcommand{\Hh}{\theta}

\newcommand{\w}{\omega}
\newcommand{\ww}[1]{\w_{\mathrm{#1}}}

\newcommand{\Gm}{\Gamma}
\newcommand{\N}{N_{i}}
\newcommand{\Com}{C}

\newcommand{\z}{\hat{\mathcal{Z}}}

\newcommand{\wc}{\omega_{c}}
\newcommand{\dwc}{\Delta\wc}
\newcommand{\kk}{\kappa}
\newcommand{\Dca}{\Delta_{\mathrm{ca}}}
\newcommand{\go}{g_{o}}

\newcommand{\n}{\bar{\nu}}
\newcommand{\nn}{\bar{n}}
\newcommand{\nb}{\dot{\nn}^{(b)}}
\newcommand{\nr}{\dot{\nn}^{(r)}}
\newcommand{\nnPh}{\nn^{\Ph}}
\newcommand{\nuth}{\bar{\nu}^{\mathrm{(th)}}}
\newcommand{\nuba}{\bar{\nu}^{\mathrm{(ba)}}}

\newcommand{\Plo}{P_{\mathrm{LO}}}
\newcommand{\hf}{\hbar\omega_{0}}
\newcommand{\edet}{\epsilon_{\mathrm{det}}}
\newcommand{\Ss}[1]{S_{\mathrm{#1}}}
\newcommand{\Ph}{(\pi/2)}

\newcommand{\Ut}{U}
\newcommand{\U}[1]{\Ut_{\mathrm{#1}}}
\newcommand{\kt}[1]{k_{\mathrm{#1}}}
\newcommand{\kp}{k_{p}}

\newcommand{\s}[1]{\sigma_{\mathrm{#1}}}

\newcommand{\za}{z_{a}}
\newcommand{\La}{a}

\begin{document}

\title{Optical read-out of the quantum motion of an array of atoms-based mechanical oscillators}



\author{Thierry Botter$^{1}$}
\email{tbotter@berkeley.edu}
\author{Daniel W. C. Brooks$^{1}$, Sydney Schreppler$^{1}$, Nathan Brahms$^{1}$}
\author{Dan M. Stamper-Kurn$^{1,2}$}
\email{dmsk@berkeley.edu}
\affiliation{
$^1$Department of Physics, University of California, Berkeley, CA, 94720, USA \\
$^2$Materials Sciences Division, Lawrence Berkeley National Laboratory, Berkeley, CA, 94720, USA}

\date{\today}

\begin{abstract}
We create an ultracold-atoms-based cavity optomechanical system in which as many as six distinguishable mechanical oscillators are prepared, and optically detected, near their ground states of motion. We demonstrate that the motional state of one oscillator can be selectively addressed while preserving neighboring oscillators near their ground states to better than 95\% per excitation quantum. We also show that our system offers nanometer-scale spatial resolution of each mechanical element via optomechanical imaging. This technique enables \emph{in-situ}, parallel sensing of potential landscapes, a capability relevant to active research areas of atomic physics and force-field detection in optomechanics.
\end{abstract}

\maketitle
Cavity-enhanced interactions between light and the motion of a mechanical element form a new structure for quantum-sensitive displacement measurements \cite{Teufel2009, Anetsberger2009}, the generation of non-classical light \cite{Brooks2012}, the storage \cite{Weis2010, Safavi2011}, control \cite{Rosenberg2009,VanThourhout2010} and transduction \cite{OConnell2010, Chang2011} of quantum information, and the optical cooling of an oscillator to its motional ground state \cite{Teufel2011, Chan2011}. To date, research efforts have focused almost exclusively on coupling a single mechanical element to light.

Cavity optomechanics with multiple moveable elements promises new possibilities, including optomechanical networks for quantum information processing \cite{Chang2011, Schmidt2012, Stannigel2012}, the entanglement of macroscopic objects \cite{Mancini2002, Pinard2005}, and experimental studies of light-mediated oscillator-oscillator interactions \cite{Ludwig2012}. Recently, researchers have begun exploring multi-oscillator optomechanics, employing pairs of nanomechanical oscillators \cite{Jiang2009, Lin2010, Zhang2011, Massel2012}. These studies have explored coupling of the mechanical modes via a shared evanescent optical field or, in the case of Ref.~\cite{Massel2012}, a shared microwave-frequency resonator. Certain classical effects were demonstrated, including the synchronization of motion, and the hybridization of modes into optically bright and dark states.

In this work, we extend these results by observing two distinctly quantum-mechanical features of an atoms-based realization of cavity optomechanics, in which several distinguishable mechanical elements, prepared near their motional ground states, are arrayed within the same optical resonator. First, we demonstrate that the near-ground-state motion of as many as six oscillators can be simultaneously measured with quantum-limited sensitivity. Second, we show that the motional state of a chosen oscillator can be changed without disturbing the near-quantum motion of neighboring oscillators.

Previous experiments investigating cavity optomechanics through the motion of atomic ensembles have used gases trapped within a single-color standing wave of light, formed by driving a Fabry-P\'erot cavity on one of its TEM$_{00}$ resonances \cite{Brennecke2008, Purdy2010, Schleier2011}. In such a potential, atoms at every lattice site oscillate at the same mechanical frequency. This mechanical degeneracy allows the motion of the entire array of trapped ensembles to be treated as that of a single mechanical element.

Here, instead, two optical trap beams of distinct colors, labeled A and B, each resonant with a separate TEM$_{00}$ mode, are injected into a Fabry-P\'erot cavity. Neglecting the variation of the cavity-field intensity in the directions transverse to the cavity axis, the resulting one-dimensional optical potential is of the form 
\begin{equation}
\Ut = \U{A}\times\mathrm{sin}^{2}(\kt{A}z)+\U{B}\times\mathrm{sin}^{2}(\kt{B}z),
\end{equation}
where $U$ and $k$ are the optical potential trap depth and wavenumber, respectively, and $z$ refers to the spatial location. In this superlattice, the potential curvature and, hence, the mechanical oscillation frequency of trapped atoms vary among neighboring lattice sites. Upon distributing atoms among several adjacent lattice sites, the collective center-of-mass vibrations of the population at each occupied site represent distinct mechanical elements that can interact optomechanically with a cavity-probe field.

\begin{figure}
	\includegraphics [width = 3.3in] {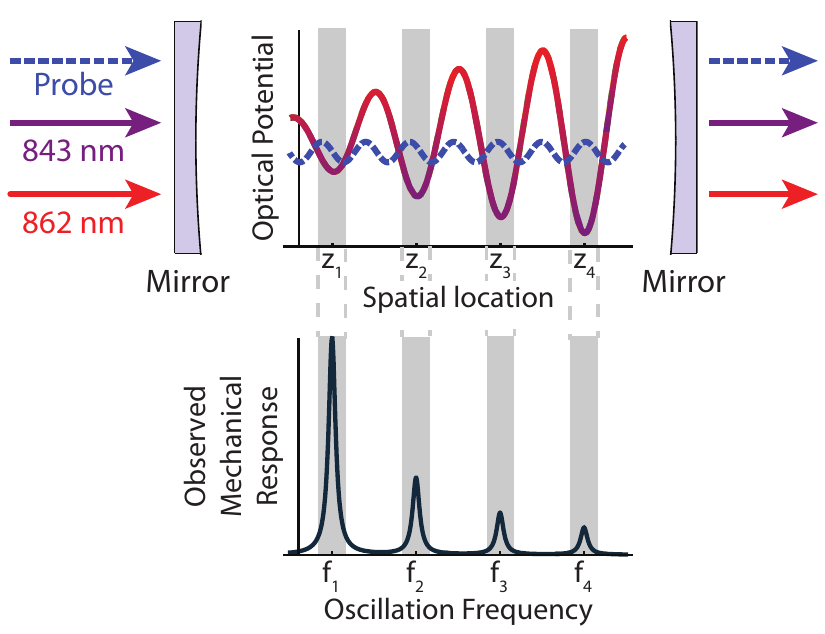}
	\caption{Experiment schematic: Superlattice potential. Three light beams are injected into an optical cavity (mirrors - grey): the probe (dashed blue), and lattice beams A (862 nm - solid red) and B (843 nm - solid purple). Atoms inserted into the cavity are trapped in potential minima of the superlattice formed by spatial beats between each lattice beam's standing-wave optical potential (red-purple gradient). The dissimilar wavelengths of lattice beams A and B causes atoms at neighboring lattice sites to resonate at distinct frequencies, which is detected via optomechanical interactions with the probe's standing-wave intensity (intracavity dashed blue). Since each potential minimum overlaps with a different probe intensity gradient, the observed response varies in amplitude among neighboring lattice sites.}  
	\label{fig:Cavity}
\end{figure}

In our experiment, a microfabricated atom chip produces an ultracold gas of $^{87}$Rb atoms that is magnetically trapped within the mode volume of a Fabry-P\'erot optical cavity of half-linewidth $\kk=2\pi\times1.82$ MHz \cite{Purdy2010}. The position of the atomic gas along the cavity axis, $\za$, is controlled by varying the magnetic trap's parameters. Tunable radio-frequency-induced evaporation allows us to reduce the atom number and also the spatial extent of the magnetically trapped atoms by lowering their temperature. The remaining atoms are then transferred from the magnetic trap to the optical superlattice formed by lattice beams A (862 nm) and B (843 nm) (Figure \ref{fig:Cavity}). The lattice depths $\U{A,B}$ are chosen so that the mechanical resonance frequencies for each of the single-color standing waves that form the superlattice are given as $\ww{A,B} = \sqrt{2\kt{A,B}\U{A,B}/m} = 2\pi\times\left(127\,\mathrm{kHz}, 128\,\mathrm{kHz}\right)$, where $m$ is the atomic mass (Supplementary Information, Figure S.1).

Information about the resulting array of oscillators is then extracted via optomechanical interactions with a third light beam, the probe beam, resonant with a separate TEM$_{00}$ cavity mode and detuned $\Dca = -40$ GHz from the atomic D2 transition (780 nm). At this detuning, atoms act primarily as a refractive medium with a single-atom cavity-QED coupling rate of $\go = 2\pi\times13.1$ MHz. The displacement, $\z_{i}$, of each collective atomic oscillator from its trap center, $z_{i}$, consequently modulates the cavity resonance frequency, $\wc$, and hence the probe's phase, by an amount
\begin{equation}\label{eq:dwc}
\dwc = \z_{i}\times\frac{\N\kp\go^{2}}{\Dca}\times\mathrm{sin}\!\left(2\kp\,z_{i}\right),
\end{equation}
where $\kp$ is the probe wavenumber, $\N$ is the number of atoms in each collective oscillator, and $\mathrm{sin}\!\left(2\kp\,z_{i}\right)$ indicates the local probe-intensity gradient \cite{Brahms2012, Brennecke2008}. The cavity's optical emission spectrum thus provides the motional spectrum of the optically sensed oscillators. Probe light transmitted through the cavity is detected using a balanced heterodyne receiver with an overall efficiency of $0.13 \pm 0.02$ for detecting cavity photons. Atoms are then released from the optical trap and the cavity probe light is again measured. This second measurement is subtracted from the first to identify and eliminate technical noise. To avoid excessive probe-induced heating of the atomic gas, data recordings are limited to 10 ms.

\begin{figure*}
	\includegraphics[width = 7.1 in] {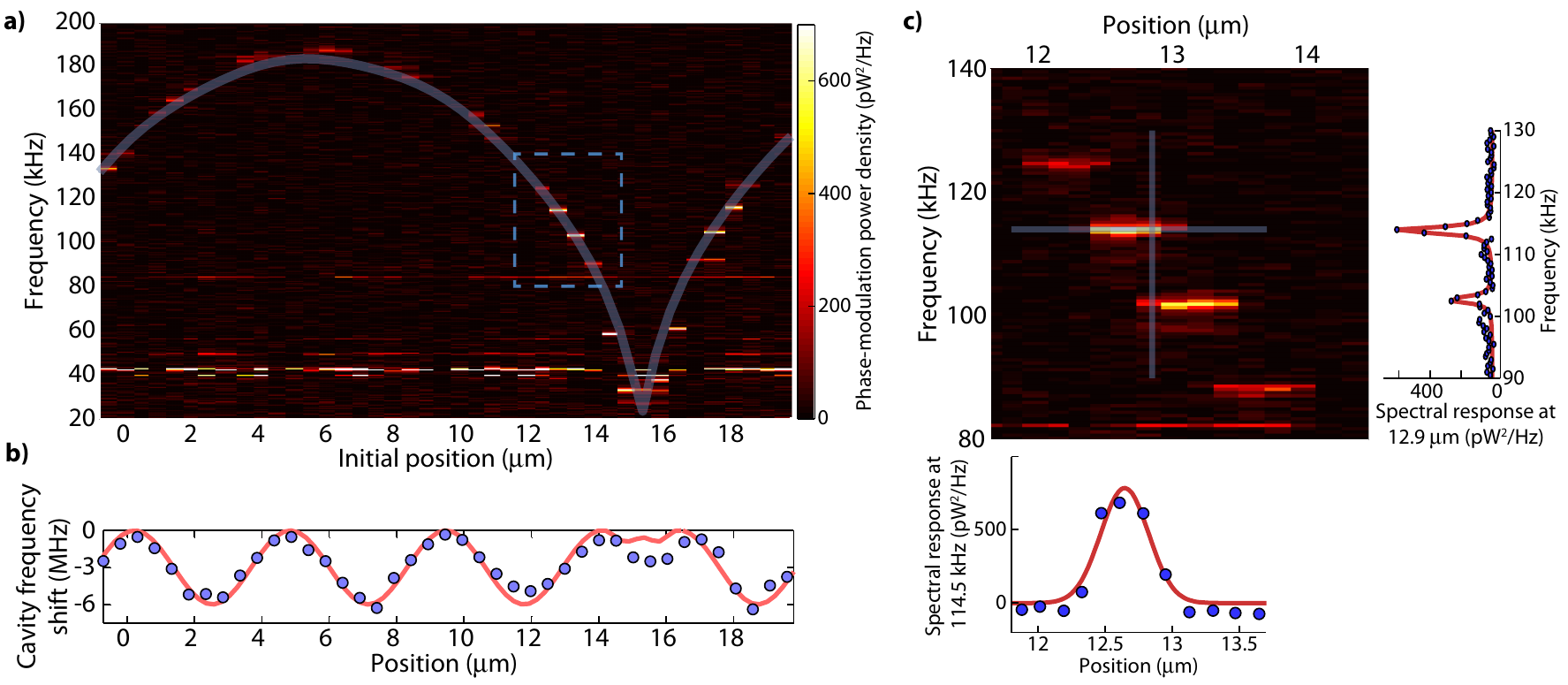}
	\caption{Distribution of mechanical resonances along the superlattice. (\textbf{a}) Phase-modulation power density as a function of frequency and loading position of the initial atomic sample. Discrete mechanical responses appear as bright pixels. They closely follow the transparent gray curve, which interpolates between the expected discrete mechanical resonances. Residual technical noise appears as horizontal bands in the figure (\emph{e.g.} near 40 kHz). (\textbf{b}) Measured shifts (purple dots) in the probe's cavity resonance are due to dispersive probe-atom coupling at each recorded superlattice position, overlaid with the modeled distribution of cavity frequency shifts (red solid line). Regions of maximum gradient are correlated with the brightest responses in (a). (\textbf{c}) Mechanical responses measured with finer steps between loading positions (within dashed box in (a)). Measured spectral responses across a selected frequency and position (transparent gray lines) are shown adjacent to the central figure (data - purple dots; fits - red solid lines). Fits yield a spatial and spectral resolution of  $\s{tot} = 0.42 \pm 0.02\, \mu$m and $\Gm = 2\pi\times(1.2 \pm 0.1)$ kHz, respectively, for the oscillator at frequency $\w=2\pi\times114.5$ kHz.
	 }
	\label{fig:Spider}
\end{figure*}

We begin by characterizing the distribution of potential minima of the optical superlattice.  In a single experimental cycle, a small cloud of $\sim 1000$ atoms is loaded into the superlattice at an initial position $\za$ chosen within a range spanning a full superlattice period $\pi / (\kt{A} - \kt{B}) = 19.4 \, \mu\mbox{m}$.  The cavity is then probed at a light intensity corresponding to an average intracavity photon number of 2.9.  From the recorded optical heterodyne signal, we extract the spectrum of motional sidebands imparted by the atomic motion, characterizing the mechanical structure of the intracavity atomic ensemble.

These phase-modulation spectra, shown in Figure \ref{fig:Spider}a, display a set of discrete mechanical resonances generated by atoms trapped at different locations within the superlattice.  Indeed, the relation between $\za$ and the observed mechanical resonances provides a detailed spatial mapping of the superlattice potential minima, one which agrees closely with the predicted form of the potential. The motional sidebands also vary strongly in their strength owing to the site-to-site variation in the linear optomechanical coupling strength (Eq.~\ref{eq:dwc}). Indeed, the probe-light intensity at each superlattice trap site can be inferred from the mean shift of the cavity resonance frequency by atoms loaded into that site (Figure \ref{fig:Spider}b).  As expected, strong motional sidebands are observed for atoms placed where this intensity shows large gradients.

The precision with which the center location of each potential minimum can be imaged depends on the spacing between adjacent lattice sites, $\La\sim420$ nm, the initial atomic ensemble's spatial full-width at half maximum (FWHM), $\s{w}$, and the precision with which $\za$ is controlled, $\s{\za}$. For infinitely narrow initial atomic distributions with arbitrarily precise initial location, atoms inserted within $\pm\La/2$ of a potential minimum will be trapped at the same location, yielding an effective spatial resolution of $\La$. For initial atomic ensembles of FWHM $\s{tot}\gg\La$, due to either large $\s{w}$ or randomly fluctuating $\za$, each oscillator's center location will be measured with a precision of $\s{tot}=\sqrt{\s{w}^2+\s{\za}^2}$. For the region considered in Figure \ref{fig:Spider}c, the spatial resolution limit is 420 nm.

\begin{figure}
	\includegraphics[width = 3.3 in] {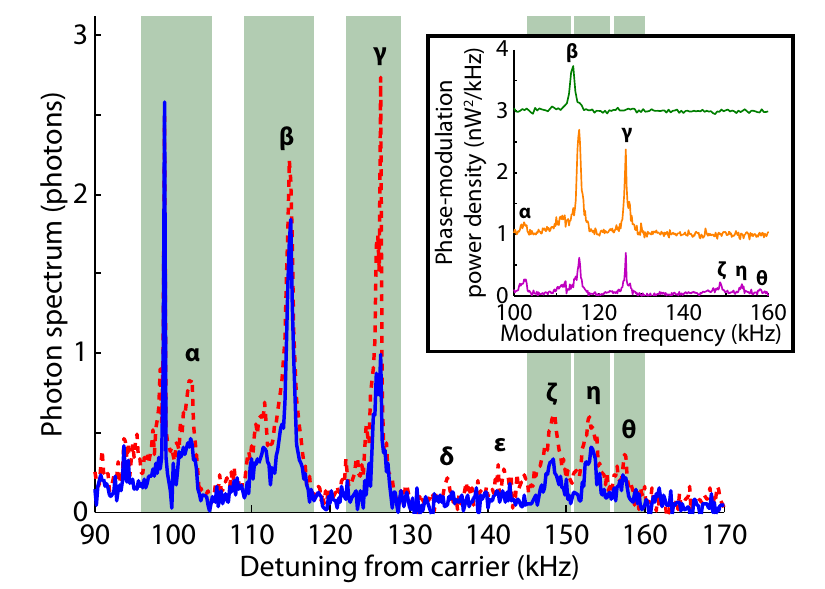}
	\caption{Stokes (dashed red) and anti-Stokes (solid blue) sideband spectrum of light exiting the cavity as a function of detuning from the probe carrier frequency. Atoms are loaded into eight neighboring lattice sites, labeled sequentially by position (Greek letters). Only six motional sidebands pairs are visible because two oscillators ($\Dd$ and $\Ee$) are positioned where the linear optomechanical response is minimal. Phonon occupations are obtained from the integrated Stokes and anti-Stokes fluorescence in narrow frequency bands around each mechanical resonance (light green): $\nu_\Aa = 1.8 \pm 0.2$, $\nu_\Bb = 2.2 \pm 0.2$, $\nu_\Cc = 1.0 \pm 0.1$, $\nu_\Ff = 0.9 \pm 0.1$, $\nu_\Gg = 1.3 \pm 0.2$, and $\nu_\Hh = 1.8 \pm 0.4$. The narrow response at 99 kHz is an optomechanically amplified technical noise spike on the probe. (\textbf{Inset}) Demonstration of experimental control over the number of constructed mechanical elements. Initial atomic ensembles of narrow (green, top), medium (orange, middle) and wide (purple, bottom) spatial extent are prepared to produce arrays of 1, 3 and 6 oscillators, respectively. Traces are offset for clarity.}
	\label{fig:6Osc}
\end{figure}

Having characterized the optical superlattice, we construct an array of distinguishable oscillators by inserting a spatially broad atomic ensemble ($\s{tot}\sim3.1\,\mu$m) at location 11.7$\mu$m (see Figure \ref{fig:Spider}a,c), where the frequency difference between neighboring mechanical resonances exceeds the mechanical resonance frequency widths, $\Gm$ (FWHM). Atoms are transfered to the many lattice sites overlapping with the initial ensemble, resulting in several distinguishable mechanical elements composed of 400 to 800 atoms each. We record the distribution of motional sidebands imprinted on the cavity-resonant probe photon spectrum, $\nn(\w)$ (Supplemental Information, section S.1). As many as six distinct oscillators are observed (Figure \ref{fig:6Osc}). In fact, for these data, atoms occupy at least eight neighboring sites of the superlattice, but two sites are not detected due to minimal linear optomechanical coupling at their locations.

We demonstrate the quantum nature of these oscillators using a parallel measurement of their Stokes emission asymmetry. Indeed, the asymmetry between the rate of red-detuned (Stokes), $\nr_{i}$, and blue-detuned (anti-Stokes), $\nb_{i}$, scattered photons, near the mechanical frequency of element $i$, reveals that mechanical element to be near its quantum mechanical ground state. That is, when the cavity is probed on resonance, this asymmetry quantifies the mean phonon occupation to be $\n_{i} = \nb_{i}/(\nr_{i}-\nb_{i})$.  Mean energy inputs from the atoms' average thermal occupation ($\nuth_{i}$) and from optomechanical interactions with the probe ($\nuba_{i}$), in equilibrium with mean energy decay via mechanical damping, sum to set $\n_{i}$: $\n_{i} = \nuth_{i}+\nuba_{i}$. Measured occupation numbers in our six-oscillator array range from 0.9 to 2.2 (Figure \ref{fig:6Osc}). These occupation numbers include the perturbative action of the probe and hence constitute upper bounds on each oscillator's native motional state, $\nuth_{i}$. As highlighted by the inset of Figure \ref{fig:6Osc}, by varying the spatial extent of the initial atomic ensemble, the number of distinct mechanical elements can be controllably tuned from one to six.

\begin{figure}
	\includegraphics[width = 3.3 in] {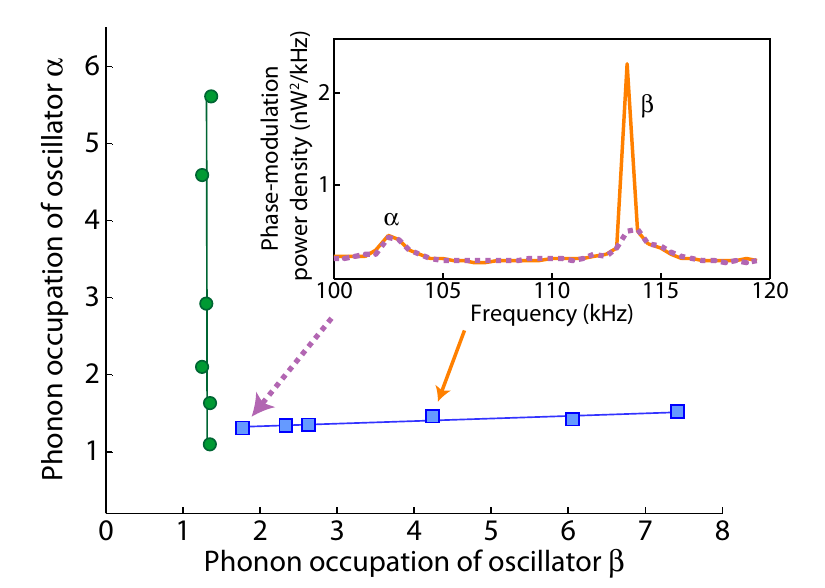}
	\caption{Selective addressing of a single intracavity oscillator. Energy is deposited into oscillator $\Aa$ (green dots) and $\Bb$ (blue squares), respectively, by coherently modulating the intensity of lattice beam B. The points correspond to the measured phonon occupation under different modulation strengths of $\U{B}$. The lines are fits to the data. From the slopes and their uncertainties, the phonon occupation of the undriven site increases by less than 0.04 (blue) and 0.05 (green), within a 97\% confidence interval, for every phonon added to the driven site. (\textbf{Inset}) Measured phase-modulation power density for two distinct drives at $\w_{\Bb}$ (purple dashed, orange solid). Arrows point to the corresponding data points.}
	\label{fig:Addressability}
\end{figure}

In addition, we demonstrate that one targeted oscillator can be driven coherently while preserving the near-quantum motion of neighboring oscillators to better than 95\% per added quantum. Oscillators $\Aa$ and $\Bb$ (Figure \ref{fig:6Osc}) are prepared with $\sim$ 750 atoms each. We apply a nearly common force modulation to both oscillators by temporarily modulating $\U{B}$, but we tune the modulation frequency to coincide with one targeted oscillator's mechanical resonance frequency. Each oscillator's squared displacement, and hence its mechanical energy, is mapped onto the probe's phase-modulation power density, which we record at the cavity output (Supplemental Information, section S.2).

Figure \ref{fig:Addressability} shows the measured simultaneous mean phonon occupation of both oscillators. The plotted occupations include energy provided by the drive in addition to the thermal and probe-induced phonon occupation. Linear fits to both series of drives indicate a base occupation of $\nuth_{\Aa, \Bb}+\nuba_{\Aa, \Bb}=1.3$. Moreover, for each quantum of collective motion added to oscillator $\Aa$ or $\Bb$, fewer than 0.05 and 0.04 phonons are added to the other oscillator, respectively. Each oscillator is therefore isolated from its neighbor by at least 95\% per added mechanical quantum. Our findings can be reframed in the context of quantum information processing: each oscillator can mechanically store information quanta for a time $1/\Gm\sim100\mu$s, information that can be retrieved via optomechanical interactions. Multi-oscillator optomechanical systems thus hold potential as quantum registers.

An important application of cavity optomechanics is precise force sensing \cite{Mamin2001}. Having an array of mechanically distinct oscillators allows one to make such force measurements simultaneously at many spatial locations, using frequency multiplexing to read out the array of force sensors with a single cavity-optical output. This can serve as a means to measure minute and sharply varying force fields, such as short-range gravitational forces and Casimir-Polder forces. Additionally, the use of optomechanics to characterize our superlattice potential (Figure \ref{fig:Spider}),  a method which could be termed  ``mechanical resonance imaging,'' can be broadly applied to directly measure any potential landscape. Sub-wavelength-scale potential variations could be resolved given our system's nanometer-scale control over the initial atomic cloud's size and location. This technique has direct relevance to the many-body physics of quantum interactions \cite{Lye2005}, Anderson localization \cite{Jendrzejewski2012, Kondov2011}, and studies of particle dynamics in different optical superlattice structures \cite{Endres2011}. 

This work was supported by the AFSOR and NSF. S.S. was supported by the Department of Defense (DoD) through the National Defense Science \& Engineering Graduate Fellowship (NDSEG) Program.

%

\newpage

\appendix

\makeatletter 
\renewcommand{\thefigure}{S\@arabic\c@figure}
\makeatother
\makeatletter 
\renewcommand{\theequation}{S\@arabic\c@equation}
\makeatother
\setcounter{figure}{0}
\setcounter{equation}{0}

\section*{Supplemental Information}

\begin{figure*}[!t]
	\includegraphics[width = 7.1 in] {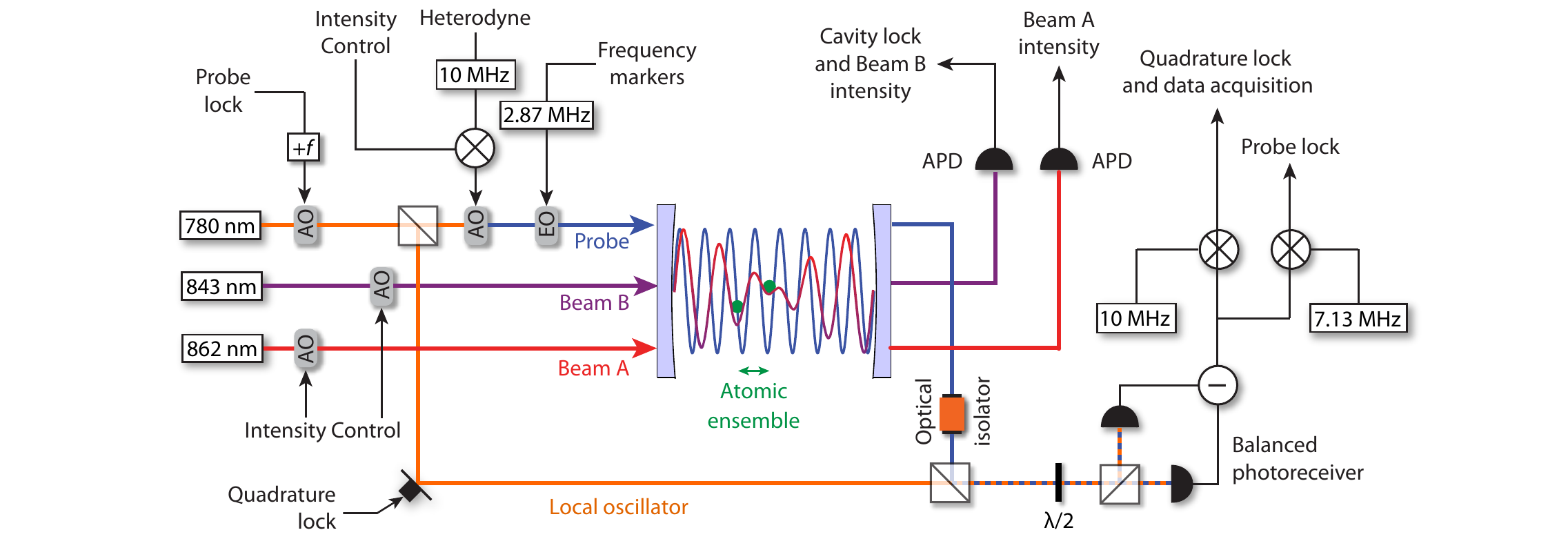}
	\caption{Schematic of experimental optical cavity. Lattice beams A (red) and B (purple), and the probe (blue) are independently monitored in transmission of the optical cavity. By adjusting their measured transmission to a desired level, $\U{A}$, $\U{B}$ and the intracavity probe intensity can be independently controlled. Lattice beam B's transmission signal is also used to feedback and lock the cavity's length. A milliwatt of 780-nm light circumvents the cavity and is used as the local oscillator (orange) for balanced heterodyne detection of the pico-watt-level probe. Recorded probe information is used to lock the local oscillator path length (quadrature lock), to control the probe-cavity detuning (probe lock) and for data analysis. Probe frequency control is accomplished by placing weak-intensity frequency markers on the probe and feeding back based on the lower frequency marker. Acronyms: AO = acousto-optical modulator, EO = electro-optical modulator, APD = avalanche photo-diode.}
	\label{fig:Apparatus}
\end{figure*}

\section*{S.1  Phonon Occupation - Probe Photon Spectrum}
For the case of an array of mechanically distinct oscillators probed on cavity resonance, the intracavity probe photon spectrum is given by 
\begin{eqnarray}\label{eq:n}
\nn(\w)& = & \sum_{i}\frac{\Com_{i}}{2}\frac{\kk^2}{\kk^2+\w^2} \\
&  & \times\left[\frac{\Gm_{i}^2\n_{i}}{(\w-\w_{i})^2+(\Gm_{i}/2)^2}+\frac{\Gm_{i}^2(\n_{i}+1)}{(\w+\w_{i})^2+(\Gm_{i}/2)^2}\right], \nonumber
\end{eqnarray}
where $\Gm_{i}$ and $\w_{i}$ symbolize the mechanical full-linewidth and oscillation frequency of oscillator $i$, respectively. The term $\Com_{i}$ represents the optomechanical cooperativity of oscillator $i$. The cooperativity parametrizes the rate of information exchange between the probe and the oscillator relative to the rate of information decay by optical and mechanical damping. For our experiment, $\Com$ was near unity for each oscillator. In the unresolved-sideband limit ($\kk \gg \w_{i}$), the phonon occupation of oscillator $i$ is related to the integrated photon number under the Stokes, $\nr_{i}$, and anti-Stokes, $\nb_{i}$, mechanical sidebands:
\begin{eqnarray}
\nr_{i} &=& \pi\Com_{i}\Gm_{i}(\n_{i}+1), \\
\nb_{i} &=& \pi\Com_{i}\Gm_{i}\n_{i}, \\
\n_{i} &=& \frac{\nb_{i}}{\nr_{i}-\nb_{i}}. \label{eq:nSB}
\end{eqnarray}

We extract $\nn(\w)$ from the power spectral density recorded by our heterodyne photodetector, $\Ss{het}(\w)$, based on the relation
\begin{equation}\label{eq:Shet}
\Ss{het}(\w) = \Ss{SN}\left[1+\edet(\nn(\w)+\nn_{0}(\w))/2\right],
\end{equation}
where $\Ss{SN} = \Plo\hf$ is the mean shot-noise spectrum, with $\Plo$ the detected local-oscillator optical power and $\hf$ the probe photon energy, $\edet=0.13 \pm 0.02$ is the overall probe cavity photon detection efficiency, and $\nn_{0}(\w)$ is the spectrum of technical noise present on the probe. Since our probe is observed to be nearly shot-noise limited at frequencies near the mechanical resonances, we make the approximation $\nn_{0}(\w) \simeq 0$. We integrate $\nn(\w)$ near each observed Stokes and anti-Stokes mechanical sidebands separately to obtain $\nr_{i}$ and $\nb_{i}$, respectively. The phonon occupation of each oscillator $i$ is then calculated from Eq.~\ref{eq:nSB}.

\section*{S.2  Phonon Occupation - Probe Phase-Modulation Power}
The phonon occupation of oscillator $\Aa$ and $\Bb$ under applied force modulations is determined from the probe's phase spectrum, $\nnPh(\w)$, which is obtained from the recorded phase-modulation power density, $\Ss{het}^{\Ph}(\w)$:
\begin{eqnarray}\label{eq:Sphase}
\Ss{het}^{\Ph}(\w) &=& \Ss{SN}^{\Ph} \\
& & \times\left[1+\frac{\edet}{2}\left(\nnPh(\w)+\nnPh_{0}(\w)\right)\right], \nonumber
\end{eqnarray}
where $\Ss{SN}^{\Ph} = \Plo\hf/2$ is the mean phase-modulation shot-noise spectrum recorded, and $\nnPh_{0}(\w)$ is the spectrum of phase-modulation technical noise on the intracavity probe field. As before, we observe the probe spectrum to be shot noise dominated near mechanical sidebands, and consequently approximate $\nnPh_{0}(\w)$ to be 0. For a cavity-resonant probe, the mechanical motion modulates only the probe's phase, and hence both the Stokes and anti-Stokes motional sidebands are mapped onto $\nnPh(\w)$: 
\begin{equation}\label{eq:nPh}
\nnPh(\w) = \sum_{i\epsilon\{\Aa,\Bb\}}\frac{\kk^2}{\kk^2+\w^2}\frac{\Com_{i}\Gm_{i}^2(2\n_{i}+1)}{(\w-\w_{i})^2+(\Gm_{i}/2)^2}
\end{equation}
where $\Gm_{i}$, $\w_{i}$, $\Com_{i}$ and $\n_{i}$ are the mechanical linewidth, mechanical resonance frequency, optomechanical cooperativity and mean phonon occupation of oscillator $i$, respectively.

We experimentally determine parameters relevant for establishing $\n_{i}$ in the presence of an applied force. The thermal phonon occupation of both oscillators, $\nuth \sim \nuth_{\Aa} \sim \nuth_{\Bb}$, is first obtained from time-of-flight thermometry of the atomic gas \cite{Brahms2012} in the absence of a probe and an applied drive. Mechanical properties are then determined by probing oscillators $\Aa$ and $\Bb$ in the absence of a drive. Fits of the observed motional sidebands in $\nnPh(\w)$ are used to extract $\Gm_{i}$, $\w_{i}$ and $\Com_{i}$, where $i = \Aa,\Bb$. Each oscillator's cooperativity is derived from the peak height of its motional sideband:
\begin{equation}\label{eq:C}
\nnPh_{i}(\w_{i}) = 4\Com_{i}(2\nuth_{i}+\Com_{i}+1), 
\end{equation}
which includes the unresolved-sideband approximation ($\kk \gg \w_{i}$) and the probe's radiation-pressure shot noise backaction on the mean phonon occupation ($\nuba_{i} = \Com_{i}/2$)\cite{Brahms2012}.

\end{document}